\documentclass[twocolumn,aps,prl,showpacs,amsmath,amssymb]{revtex4}
\usepackage{graphicx,dcolumn,bm,color,framed}
\begin{document}

\title{Dynamical Quantum Anomalous Hall Effect in Strong Optical Fields}
%the Intense Optical Field Regime}

\author{Woo-Ram Lee}
\author{Wang-Kong Tse}
%\email{wktse@ua.edu}
\affiliation{
Department of Physics and Astronomy, The University of Alabama, Alabama 35487, USA\\
Center for Materials for Information Technology, The University of Alabama, Alabama 35401, USA}
%\affiliation{Department of Physics and Astronomy, Center for Materials for Information Technology, The University of Alabama, Alabama 35487, USA}

\date{\today}

\begin{abstract}
Topological insulators (TIs) are characterized by the quantum anomalous Hall effect (QAHE) on the topological surface states under time-reversal symmetry breaking. Motivated by recent experiments on the magneto-optical effects induced by the QAHE, we develop a theory for the dynamical Hall conductivity for subgap optical frequency and intense optical fields using the Keldysh-Floquet Green's function formalism. 
%While this effect has been studied theoretically and experimentally under magneto-optical using weak linearly-polarized light as a probe field, the influence of strong optical pump fields on QAHE remains largely not understood. Using the Keldysh-Floquet Green's function formalism, we develop a theory for the dynamical Hall conductivity for subgap optical frequency and intense optical fields. 
Our theory reveals a nonlinear regime in which the Hall conductivity remains close to $e^2/2h$ at low frequencies. At higher optical fields, we find that the subsequent collapse of the half quantization is accompanied by coherent oscillations of the dynamical Hall conductivity as a function of field strength, triggered by the formation of Floquet subbands and the concomitant inter-subband transitions. 
%We apply our theory in particular to the adiabatic, low-frequency regime, and study the robustness and collapse of the Hall quantization under strong optical fields. Our results reveal that the nonlinear dynamical Hall conductivity exhibits coherent oscillations as a function of the field strength, which is triggered by the formation of Floquet subbands and the concomitant inter-subband transitions. 
%coherent oscillation of the dynamical Hall conductivity as a function of the field strength
%the coherent oscillation of the dynamical Hall conductivity depending on the incident optical field
%a strong nonlinear dependence of the dynamical Hall conductivity on the incident optical field, which is triggered by the formation of Floquet subbands and the concomitant inter-subband transitions. 
%between them.
\end{abstract}

\maketitle

%%%%%%%%%%%%%%%%%%%%%%%%
%                        Introduction                         %
%%%%%%%%%%%%%%%%%%%%%%%%

\emph{Introduction}.--- Topological quantum phases of matter are one of the most 
%dominant 
intriguing paradigms in contemporary condensed matter physics~\cite{Hasan10, Qi11}. Recently, intensive researches have been focused on the interplay between topological order and dynamics. This is well exemplified by the dynamical synthesis and manipulation of topological quantum phases, which include Floquet Chern insulators~\cite{Oka09}, Floquet 
%topological insulators
TIs~\cite{Lindner11}, Floquet Majorana fermions~\cite{Jiang11, Liu12}, and Floquet Weyl semimetals~\cite{Wang14, Chan16}. 
%On the other hand, the nonlinear dynamical response of noncentrosymmetric crystals possibly opens a new route towards the design of optoelectronic devices~\cite{Morimoto16, Kim16}. 

Topological Hall quantization is a hallmark signature of two-dimensional (2D) quantum anomalous Hall insulators~\cite{Liu16}. The crucial ingredients to realize the quantum anomalous Hall state 
%is
are strong spin-orbit coupling and a magnetic Zeeman gap $\Delta$. With magnetic doping, these can be realized on 
%the surface of a topological insulator
the TI surface~\cite{QAHE_Yu}, single-layer or bilayer graphene \cite{QAHE_Tse1,QAHE_Tse2}, and HgTe quantum well \cite{QAHE_Liu}. 
%on the surface of a topological insulator \cite{} or a Chern insulator \cite{} with an exchange-induced magnetic gap $\Delta$. 
Under a \textit{D.C.} bias, a robust Hall plateau corresponding to the quantized Hall value has already been observed~\cite{Chang13, Chang15, Kou15} in magnetically doped 
%topological insulator
TI films. 
%In a transport measurement setup under a weak in-plane \textit{D.C.} electric field, a robust Hall plateau was observed~\cite{Chang13, Chang15, Kou15}. 
While topological transport as a property of linear response has been 
%extenstively
extensively studied under a \textit{D.C.} or an \textit{A.C.} electric field serving as a weak probe field, one is naturally led to consider whether nonlinear responses can exhibit topological properties under strong optical pump fields. In this vein, recent works have shown that the second and third order optical susceptibilities~\cite{Nagaosa1, Nagaosa2, Moore, Kim16} in topological matter may also contain topological contributions arising from a nonzero Berry curvature. An important unanswered question in the emerging topic of nonequilibrium topological response concerns the strong-field influences on the topological Hall conductivity. Theoretical predictions~\cite{Tse10, Maciejko10} have shown that the quantization of the Hall conductivity can manifest as quantization of magneto-optical Faraday and Kerr rotations when 
%topological insulator
TI thin films are illuminated by a low-frequency optical probe field, as confirmed in recent experiments \cite{Okada16, Wu16, Dziom16}. This provides a strong motivation to consider the effects on the topological Hall quantization due to a strong \textit{A.C.} driving field $E$, particularly in the low-frequency subgap regime ($\hbar\Omega \ll \Delta$), when the light-matter interaction $\sim E/\Omega$ is manifestly 
%non-perturbative. 
nonperturbative.

In this work, we develop a theory for the dynamical Hall effect in quantum anomalous Hall insulators using the Keldysh-Floquet Green's function formalism. We note that the study of topological 
states using the Floquet formalism (in Floquet 
%topological insulators,
TIs, for example) has so far been largely focused on the high-frequency, off-resonance regime; the effects of low-frequency \textit{A.C.} driving in the adiabatic regime, where the frequency is small compared to either the bandwidth or the band gap, remain not well understood. To remedy this situation, we numerically study the 
%anomalous 
Hall conductivity as a function of the optical field strength and frequency without any expansion in powers of optical field strength. This approach allows us to address the full 
%non-perturbative
nonperturbative effects of the optical field on the electronic bands and transport properties of the system, even at low frequencies. For concreteness, we take the massive Dirac model describing 
%topological insulator
TI surface states with broken time-reversal symmetry~\cite{Chen10, AndrewWray11} as our prototypical system of investigation. We expect 
%that 
the conclusions from our theory to be broadly applicable to other materials that host massive Dirac fermions such as  Chern insulators and graphene-like materials with broken spatial-inversion symmetry. 

\emph{2D massive Dirac fermions and QAHE}.--- 
We consider the quantum anomalous Hall state realized on the surface of a 
%topological insulator (TI)
TI film with broken time-reversal symmetry, \textit{e.g.}, by interfacing with a  magnetic substrate or by doping with magnetic adatoms~\cite{Hasan10, Qi11, Liu16}. The film is sufficiently thick so that tunneling between the top and bottom surface states can be ignored \cite{Lu10}. Focusing on a single surface, a low-energy electron at the $\Gamma$ point in the Brillouin zone is described by the 2D massive Dirac Hamiltonian $\mathcal{H} = \sum_{{\bf k}} \psi_{\bf k}^\dag \mathcal{H}_{{\bf k}} \psi_{\bf k}$ with $\mathcal{H}_{{\bf k}} = {\bf d}_{{\bf k}} \cdot {\boldsymbol \tau}$. Here, 
%$\psi_{\bf k} = (c_{{\bf k} A}, c_{{\bf k} B})^T$ 
$\psi_{\bf k} = (c_{{\bf k} \uparrow}, c_{{\bf k} \downarrow})^T$ 
with $c_{{\bf k}\alpha}$ 
%is 
%{\color{red} being} 
the electron annihilation operator with momentum $\hbar{\bf k}$ and 
%pseudospin $\alpha(=A,B)$,
spin $\alpha(=\uparrow,\downarrow)$, 
${\boldsymbol \tau} = (\tau_x, \tau_y, \tau_z)$ 
consists of the Pauli matrices, and %the ${\bf d}_{\bf k}$ vector has components 
${\bf d}_{\bf k} = (v \hbar k_x, v \hbar k_y, m_0)$ with $v$ being the Dirac 
%Fermi 
velocity. In the presence of the Dirac mass $m_0$, or equivalently, the band gap $\Delta (= 2m_0)$, 
which is generated by the exchange field due to the magnetic substrate or adatoms, 
%the Zeeman field originating from magnetic dopants, 
the bulk energy spectrum for the surface state becomes insulating with the energy dispersion $\mathcal{E}_{\bf k} = |{\bf d}_{\bf k}| = [(v \hbar)^2 (k_x^2 + k_y^2) + m_0^2]^{1/2}$ for the conduction band and $-\mathcal{E}_{\bf k}$ for the valence band. In the presence of a time-independent \textit{D.C.} electric field, 2D massive Dirac fermions give rise to the QAHE consisting of a robust one-half quantization of the Hall conductance in units of $\sigma_0 = e^2/h$ due to the half-skyrmion configuration of $\hat{{\bf d}}_{\bf k}$~\cite{Liu16}: ${\sigma_{xy}}/{\sigma_0} = ({1}/{4\pi}) \int d^2{\bf k}~\hat{{\bf d}}_{\bf k} \cdot ( \partial_{k_x} \hat{{\bf d}}_{\bf k} \times \partial_{k_y} \hat{{\bf d}}_{\bf k}) = {\rm sgn}(m_0)/2$,
%
% \begin{align}
% \frac{\sigma_{xy}}{\sigma_0} 
% = \frac{1}{4\pi} \int d^2{\bf k}~\hat{{\bf d}}_{\bf k} \cdot ( \partial_{k_x} \hat{{\bf d}}_{\bf k} \times \partial_{k_y} \hat{{\bf d}}_{\bf k}) = %\frac{{\rm sgn}(m_0)}{2},
% \frac{1}{2} \frac{m_0}{|m_0|},
% \label{Quantization}
% \end{align}
where $\hat{{\bf d}}_{\bf k} = {\bf d}_{\bf k} / |{\bf d}_{\bf k}|$, and `$\textrm{sgn}$' denotes the signum function. 

% A half-integer in Eq.~\eqref{Quantization} is caused by the half-skyrmion configuration of $\hat{{\bf d}}_{\bf k}$. 
%As mentioned, the QAHE is also captured in a magneto-optical measurement setup. 
%In the following, we formulate a theory for the dynamical Hall conductivity valid for frequency values smaller than the band gap and for strong optical field strengths in a %nonperturbative fashion.  

%%%%%%%%%%%%%%%%%%%%%%%%
%               Dynamical Hall conductivity             %
%%%%%%%%%%%%%%%%%%%%%%%%

\emph{Dynamical response to the optical field}.--- 
%We consider the TI film illuminated by a linearly polarized light in the normal direction to the surface.
%We consider a TI film illuminated by a linearly polarized light in the normal direction to the surface.
We now consider linearly polarized light  illuminated in the normal direction to the surface of the TI film. Choosing the polarization direction 
%of the incident electric field in 
along the $x$ axis and the propagation direction along the $z$ axis, 
%and ignoring the component of a tiny magnetic field, 
the incident light with electric field amplitude $E$ and frequency $\Omega$ is described by the vector potential ${\bf A}(t) = - ({c E}/{\Omega}) \zeta(t) \sin(\Omega t)\hat{x}$.  
%which is connected to the electric field ${\bf E}(t) = - \frac{1}{c} \frac{\partial}{\partial t} {\bf A}(t)$. 
Here, a switching protocol is encoded in the function $\zeta(t) = e^{t / \tau_s} \Theta(-t) + \Theta(t)$, with $\tau_s$ being a switch-on time scale and $\Theta(x)$ the step function, which satisfies $\zeta \rightarrow 0$ for an equilibrium state at $t\rightarrow-\infty$ and $\zeta = 1$ for a nonequilibrium steady state (NESS) at $t\geq0$. In the semiclassical treatment of the optical field, the Peierls substitution $\hbar {\bf k} \rightarrow \hbar {\bf k} + {e} {\bf A}(t)/c$ in the original Hamiltonian leads to the time-dependent Hamiltonian $\mathcal{H}_{\bf k}(t) = \mathcal{H}_{\bf k} + \mathcal{V}(t)$, with the perturbation 
%taking the form 
of the form
$\mathcal{V}(t) = - \mathcal{V}_0 \zeta(t) \sin(\Omega t) \tau_x$ with $\mathcal{V}_0 = eEv/\Omega$ playing the role of Rabi frequency of the two-band system. The relative strength of the Rabi frequency to the photon energy defines the dimensionless coupling parameter $\lambda = \mathcal{V}_0/\hbar\Omega$, according to which the system is in a regime commonly classified as weak coupling ($\lambda \ll 1$) and strong coupling ($\lambda \gtrsim 1$)~\cite{coupling1,coupling2} in quantum optics.
%can be characterized as being in the weak coupling ($\lambda \ll 1$) and strong coupling ($\lambda \gtrsim 1$) regimes~\cite{coupling1,coupling2} in literature of quantum %optics.}

For the purpose of formulating a 
%nonperturbative 
nonperturbative theory of dynamical response, we start with a generic form of the surface electric current density, 
${\bf J}(t) = - e S^{-1} \sum_{\bf k} \langle {\bf v}_{\bf k}(t) \rangle$, where $S$ is the normalization area,
%${\bf J}(t) = - e~{\rm lim}_{S\rightarrow0}$ $S^{-1} \sum_{\bf k}
%\langle {\bf v}_{\bf k}(t) \rangle$, where $S$ is a normalization area, %unit area on the surface, 
${\bf v}_{\bf k}(t) = \psi_{\bf k}^\dag(t) \nabla_{\bf k} [\mathcal{H}_{{\bf k}}(t)/\hbar] \psi_{\bf k}(t)$ is  the single-electron velocity operator, and $\langle \mathcal{M} \rangle = {\rm Tr}(\rho_0 \mathcal{M})$ is the canonical ensemble average of the operator $\mathcal{M}$ in terms of the initial density matrix $\rho_0 = e^{-\mathcal{H}/k_{\rm B}T} / {\rm Tr}(e^{-\mathcal{H}/k_{\rm B}T})$, which depends on the switching protocol as discussed later. The time evolution of ${\bf J}(t)$ is expressed by using the lesser Green's function, $[\hbar G_{\bf k}^<(t,t)]_{\alpha\beta} = i \langle c_{{\bf k}\beta}^\dag(t) c_{{\bf k}\alpha}(t) \rangle$. In this work, if we focus on a NESS, the system recovers time translational symmetry, i.e., $\mathcal{H}(t) = \mathcal{H}(t+\tau)$ with periodicity $\tau (= 2\pi/\Omega)$, and is  thus governed by the Floquet 
%theory
theorem~\cite{Grifoni98}. By considering the Floquet mode expansion of the Green's function, $[\hbar G_{\bf k}^<(t,t)]_{\alpha\beta} = \sum_{m,n\in\mathbb{Z}} e^{-i(m-n)\Omega t}$ 
$\int_{-\hbar\Omega/2}^{\hbar\Omega/2} {d\hbar\omega} [\hat{G}_{\bf k}^<(\omega)]_{\alpha\beta;mn}/{2\pi}$, we find the surface current density $J_{\mu}(t) = \sum_{s\in\mathbb{N}} {\rm Re} \big[ \sigma_{\mu x}^{s}(E) E e^{-i s \Omega t} \big]$ $(\mu\in\{x,y\})$, 
where the $s$-th harmonics of the field-dependent dynamical longitudinal and Hall conductivities are given by
%defined by
\begin{align}
{\rm Re}[\sigma_{xx}^{s}(E)] 
& = {\rm Im}[\tilde{\sigma}_{+}^{s}(E)],
\\ 
{\rm Im}[\sigma_{yx}^{s}(E)]
& = {\rm Im}[\tilde{\sigma}_{-}^{s}(E)],
\end{align}
for the dissipative (incoherent) components, and  
\begin{align}
{\rm Re}[\sigma_{yx}^{s}(E)]
& = {\rm Re}[\tilde{\sigma}_{+}^{s}(E)],
\\
{\rm Im}[\sigma_{xx}^{s}(E)] 
& = - {\rm Re}[\tilde{\sigma}_{-}^{s}(E)],
\end{align}
for the reactive (coherent) components, and $\tilde{\sigma}_{\pm}^{s}(E)$ reads
\begin{align}
\frac{\tilde{\sigma}_{\pm}^{s}(E)}{\sigma_0}
& = - 2 \frac{E_0}{E} \left(\frac{v\hbar}{\Delta}\right)^2 \frac{1}{S} \sum_{\bf k} \sum_{n\in\mathbb{Z}} \int_{-\hbar\Omega/2}^{\hbar\Omega/2} d\hbar\omega
\nonumber\\
& ~~~ \times \Big\{ [\hat{G}_{\bf k}^<(\omega)]_{\uparrow\downarrow;n+s,n} \pm [\hat{G}_{\bf k}^<(\omega)]_{\uparrow\downarrow;n,n+s} \Big\}.
\label{GeneratingFunction}
\end{align}
Here, $E_0 = \Delta^2/(ev\hbar)$ defines a natural scale of maximum field strength in our problem at which Zener breakdown occurs~\cite{Zener34}, 
%unit for the optical field corresponding to the maximum field strength at which Zener breakdown occurs \cite{Zener_ref},  
%defines a natural unit for the optical field corresponding to the maximum field strength at which Zener breakdown occurs \cite{Zener_ref},  
%defines the unit optical field, 
and 
%the momentum sum is evaluated by 
$\sum_{\bf k} = S (2\pi v\hbar)^{-2} \int_{\Delta/2}^{\mathcal{E}_c} d\mathcal{E}_{\bf k}~\mathcal{E}_{\bf k} \int_{0}^{2\pi} d\varphi_{\bf k}$ with $\varphi_{\bf k} = \tan^{-1}(k_y / k_x)$ and $\mathcal{E}_c$ being an ultraviolet energy cutoff for the Dirac model at the level of low-energy effective theory.
%of the tight-binding model. 

%%%%%%%%%%%%%%%%%%%%%%%%
%  Green's functions in Keldysh-Floquet  space  %
%%%%%%%%%%%%%%%%%%%%%%%%

\emph{Green's function in the Keldysh-Floquet space}.--- 
Now, our main task %is boiled down 
boils down to finding the lesser Green's function $\hat{G}_{\bf k}^<$ in 
%the dynamical conductivity formula, 
Eq.~\eqref{GeneratingFunction}, 
%The remaining task can be achieved
which can be achieved using the Keldysh-Floquet Green's function formalism~\cite{Tsuji09, Lee14}. 
%In this formalism, the Keldysh contour is introduced in the time domain to guarantee the Gell-Mann and Low theorem in a generic nonequilibrium situation~\cite{Rammer86,Haug08}. 
%In this formalism, the Keldysh contour is introduced in the time domain, because quantum states in the remote past and future are not adiabatically connected to each other under non-equilibrium conditions~\cite{Rammer86,Haug08}. 
In this formalism, the Keldysh contour is introduced in the time domain, since quantum states at remote past and future on a single time-ordered branch are not adiabatically connected to each other under nonequilibrium condition~\cite{Rammer86,Haug08}. 
Especially, regarding NESS, the contour-ordered Green's function is mapped onto the Keldysh space, represented by a $2\times2$ matrix. This Keldysh space is further extended to include the Floquet space for systems satisfying the Floquet theorem.

The interacting Green's function is governed by the Dyson 
%Keldysh-Dyson 
equation, represented in the Keldysh-Floquet space as follows:
\begin{align}
& \left(
\begin{array}{cc}
\hat{G}_{\bf k}^R & \hat{G}_{\bf k}^K \\
0 & \hat{G}_{\bf k}^A
\end{array}
\right)^{-1}
= 
\left(
\begin{array}{cc}
\hat{\mathcal{G}}_{\bf k}^R & \hat{\mathcal{G}}_{\bf k}^K
\\
0 & \hat{\mathcal{G}}_{\bf k}^A
\end{array}
\right)^{-1} 
- \left(
\begin{array}{cc}
\hat{\mathcal{V}} & 0
\\
0 & \hat{\mathcal{V}}
\end{array}
\right).
\label{KeldyshDysonEq}
\end{align}
On the left-hand side of Eq.~\eqref{KeldyshDysonEq}, the retarded, advanced, and Keldysh components are defined by $[\hat{G}_{\bf k}^{\gamma}(\omega)]_{\alpha\beta;mn} = \int dt e^{i (\omega + m\Omega) t} \int dt' e^{-i (\omega + n\Omega) t'} [G_{\bf k}^{\gamma}(t, t')]_{\alpha\beta}$ ($\gamma = R,A,K$), where $[\hbar G_{\bf k}^{R,A}(t,t')]_{\alpha\beta} = \mp i \Theta(\pm t \mp t') \langle \{c_{{\bf k} \alpha}(t), c_{{\bf k} \beta}^\dag(t')\} \rangle$, $[\hbar G_{\bf k}^K(t,t')]_{\alpha\beta} = - i \langle [c_{{\bf k} \alpha}(t), c_{{\bf k} \beta}^\dag(t')] \rangle$, and $-\Omega/2 \leq \omega < \Omega/2$. In the right-hand side of Eq.~\eqref{KeldyshDysonEq}, the first term describes the initial NESS at $t=0$ after transient effects have washed out, and the second term corresponds to the perturbation defined by $(\hat{\mathcal{V}})_{mn} = 
\tau^{-1} \int_{0}^{\tau} dt e^{i (m-n) \Omega t} \mathcal{V}(t) = - i (\mathcal{V}_0/2) \tau_x (\delta_{m,n+1} - \delta_{m,n-1})$. 
%In principle, if 
Provided that the initial NESS is known, the interacting Green's function can be found by numerically inverting Eq.~\eqref{KeldyshDysonEq}. 
% the interacting Green's function can be found by numerically inverting Eq.~\eqref{KeldyshDysonEq}, only if the initial NESS is known, and the Floquet space dimension is also reasonably cut off. 
Then, the result is inserted into Eq.~\eqref{GeneratingFunction} through the relation 
%relationship 
$\hat{G}_{\bf k}^< = (\hat{G}_{\bf k}^K - \hat{G}_{\bf k}^R + \hat{G}_{\bf k}^A)/2$.

In Eq.~\eqref{KeldyshDysonEq}, the initial NESS was not specified yet.  We assume that the optical field is adiabatically switched on with the switch-on time $\tau_s$ long enough compared with other time scales in the system, so as to restore the condition of thermal equilibrium at $t=0$ \cite{Remark1}. Under this condition, the initial NESS is described by the equilibrium Green's function with the components: $\hat{\mathcal{G}}_{\bf k}^{\gamma}(\omega) = \mathcal{U}_{\bf k}~\hat{g}_{\bf k}^{\gamma}(\omega)~\mathcal{U}_{\bf k}^\dag$  ($\gamma = R,A,K$), where $[\hat{g}_{\bf k}^R(\omega)]^{-1} = \mathbb{I}_2 \otimes [(\hbar\omega + i \eta) \mathbb{I}_{\infty} + \hbar\hat{\Omega}] - \mathcal{E}_{\bf k} \tau_z \otimes \mathbb{I}_{\infty}$ and $[\hat{g}^A_{\bf k}(\omega)] = [\hat{g}^R_{\bf k}(\omega)]^\dag$. 
$\hat{g}^K_{\bf k}$, satisfying the fluctuation-dissipation relation~\cite{Haug08}, is given by  $\hat{g}^K_{\bf k}(\omega) = [\hat{g}^R_{\bf k}(\omega) - \hat{g}^A_{\bf k}(\omega)] \mathbb{I}_2 \otimes [\mathbb{I}_{\infty} - 2\hat{\mathcal{F}}(\omega)]$. 
%and $\hat{g}^K_{\bf k}(\omega) = [\hat{g}^R_{\bf k}(\omega) - \hat{g}^A_{\bf k}(\omega)] \mathbb{I}_2 \otimes [\mathbb{I}_{\infty} - 2\hat{\mathcal{F}}(\omega)]$, which satisfy %the fluctuation-dissipation relation. 
Here, $\mathbb{I}_n$ is the $n \times n$ identity matrix, $\eta$ is
the band broadening 
%level broadening 
width due to elastic scattering of electrons, $(\hat{\Omega})_{mn} = n \Omega \delta_{mn}$, and $[\hat{\mathcal{F}}(\omega)]_{mn} = f_{\rm FD}(\omega+n\Omega) \delta_{mn}$ with $f_{\rm FD}(\omega) = (e^{\hbar\omega/k_{\rm B}T} + 1)^{-1}$. 
%The unitary operator $\mathcal{U}_{\bf k}$ connects the Green's function in the spin basis with the counterpart in the band basis, in the form $\mathcal{U}_{\bf k} = \sin\vartheta_{\bf k} (\cos\varphi_{\bf k} \tau_x + \sin\varphi_{\bf k} \tau_y) + \cos\vartheta_{\bf k} \tau_z$ with $\vartheta_{\bf k} = \cos^{-1}[\Delta/(2\mathcal{E}_{\bf k})] / 2$. 
The unitary operator 
%$\mathcal{U}_{\bf k} = \sin\vartheta_{\bf k} (\cos\varphi_{\bf k} \tau_x + \sin\varphi_{\bf k} \tau_y) + \cos\vartheta_{\bf k} \tau_z$ 
%$\mathcal{U}_{\bf k} = \sin(\vartheta_{\bf k}/2) (\cos\varphi_{\bf k} \tau_x + \sin\varphi_{\bf k} \tau_y) + \cos(\vartheta_{\bf k}/2) \tau_z$ 
$\mathcal{U}_{\bf k} = [\sin(\vartheta_{\bf k}/2) (\cos\varphi_{\bf k} \tau_x + \sin\varphi_{\bf k} \tau_y) + \cos(\vartheta_{\bf k}/2) \tau_z] \otimes \mathbb{I}_{\infty}$
%transforms 
%is the unitary operator that transforms 
transforms the original spin representation of 
%the Hamiltonian 
$\mathcal{H}_{{\bf k}}$ 
%to 
into the band representation, with  
%converts between the spin and the band representations with
%the conversion factor 
%between the spin and band bases with 
%$\vartheta_{\bf k} = \cos^{-1}[\Delta/(2\mathcal{E}_{\bf k})] / 2$. 
$\vartheta_{\bf k} = \cos^{-1}[\Delta/(2\mathcal{E}_{\bf k})]$.
%(see Supplementary Material for more details).
%WK: Please check whether the above expression for $\vartheta_{\bf k}$ is correct, is the 1/2 factor outside or inside the cos^{-1}?
%
%WR: To avoid confusion, let's rescale: \varphi_{\bf k} --> \varphi_{\bf k} / 2.

%%%%%%%%%%%%%%%%%%%%%%%%
% Linear response and near-resonance regimes  %
%%%%%%%%%%%%%%%%%%%%%%%%

%\emph{Recovery of physical limits}.--- 
%\emph{Linear Response and Near-Resonance Regimes}.--- 
\emph{Linear-response and near-resonance regimes}.--- 
Our formalism provides a generic framework to investigate the dynamical response of massive Dirac electrons to strong optical fields. Before applying our theory to the full 
%non-perturbative
nonperturbative regime, here we consider in particular (i) the linear-response and (ii) the near-resonance coherent regimes, and see whether our framework reproduces established results in these well-known limits. 
%it is valuable to check if known physical limits are properly recovered. Below, we focus on (i) the linear-response regime; (ii) the near-resonance {\color{red}coherent} regime. 

\begin{figure}[t]
\centering
\includegraphics[width=0.45\textwidth]
{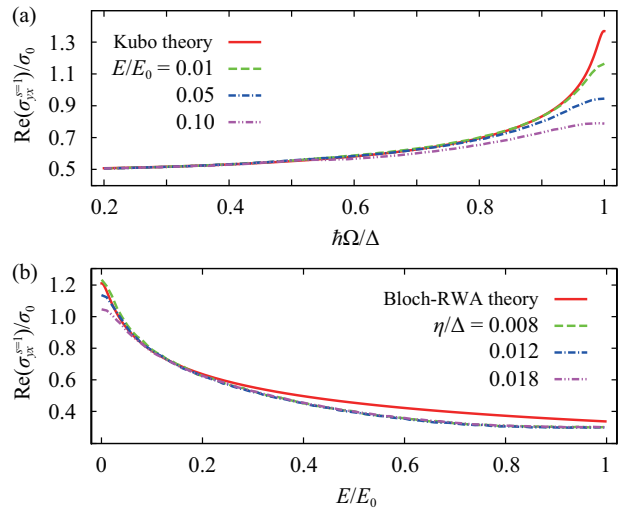} \\
\caption{
Comparison of the dynamical Hall conductivity obtained by our theory with the counterpart (a) by the Kubo theory in the 
%linear response
linear-response regime; (b) by the Bloch theory in the rotating wave approximation (RWA) near resonance. In (a), we set $\eta / \Delta = 0.01$ with varying $E$; in (b), $\hbar\Omega / \Delta = 0.99$, or equivalently the detuning $\delta / \Delta = 1 - \hbar\Omega / \Delta = 0.01$, with varying $\eta$. In both (a) and (b), we have common parameters $k_B T / \Delta = 0.02$ and $\mathcal{E}_c / \Delta = 10$.}
\label{Figure1}
\end{figure}

First, we are able to recover the linear-response result when $\mathcal{V}_0 \ll \Delta$. In this regime, Eq.~\eqref{KeldyshDysonEq} can be expanded analytically up to the linear order in $\mathcal{V}_0$, and we recover
%to derive 
the Kubo formula result for dynamical conductivity~\cite{Tse10} (see Supplementary Material for details).
%The first step is to invert the retarded component: 
%$\big[\hat{G}^{r}_{\bf k}(\omega) \big]_{\alpha\alpha;mn}^{-1} = \big[\hat{\mathcal{G}}^{r}_{\bf k}(\omega) \big]_{\alpha\alpha;mm}^{-1} \delta_{mn} +  \mathcal{V}_0 \frac{i}{2} \big\{ [\mathcal{H}_{\bf k}]_{\alpha\bar{\alpha}} \hat{g}_{{\bf k}\bar{\alpha} m}^{R0}(\omega)  + [\mathcal{H}_{\bf k}]_{\bar{\alpha}\alpha} \hat{g}_{{\bf k}\bar{\alpha} n}^{R0}(\omega)  \big\} (\delta_{m,n+1} - \delta_{m,n-1})$ $+ \mathcal{O}(\mathcal{V}_0^2)$, where $\hat{g}_{{\bf k}\alpha m}^{R0}(\omega) = \big[ \hbar(\omega + m\Omega) - [\mathcal{H}_{\bf k}]_{\alpha\alpha} + i\eta \big]^{-1}$, which 
%%is decoupled from the others in the Keldysh space, and 
%is tridiagonal in the Floquet space. For a tridiagonal matrix, an analytic inversion formula is available~\cite{Huang97}. The advanced component can be found in a similar way. The next step is to expand the lesser component, which depends on the other components in the Keldysh space. In the final step, we insert the result into Eq.~\eqref{GeneratingFunction} and are able to recover the same result for the dynamical conductivity obtained from the Kubo formula ~\cite{Tse10} (see Supplementary Material for details).
%The final step is to insert the result into Eq.~\eqref{GeneratingFunction} to reproduce the Kubo formula result for the dynamical conductivity~\cite{Tse10} (see Supplementary %Material for details). 
In Fig.~\ref{Figure1}(a), we also numerically confirm that the linear-response behavior of the dynamical Hall conductivity is recovered as the optical field strength is decreased (the reference plot of the Kubo formula result is indicated by the red solid line) \cite{Remark2}. In the low-frequency regime, we notice that the one-half Hall quantization remains robust against weak fields.

Secondly, we examine the near-resonance coherent regime, where the optical frequency is close to the band gap with the detuning $\delta (= \Delta - \hbar\Omega)$ satisfying $\Delta \gg \delta \gg \eta$. For the Dirac model, a theory based on the Bloch equation has been developed within the rotating
wave approximation (RWA)~\cite{Tse16} (see Supplementary Material for details). The result for the dynamical Hall conductivity from this theory is plotted in Fig.~\ref{Figure1}(b) as the red solid line. 
%In the weak-field ($E/E_0 \lesssim 0.2$) regime, 
For weak fields ($E/E_0 \lesssim 0.2$), as $\eta$ decreases below $\delta$ approaching the coherent regime ($\eta \to 0$), we find close agreement between the Keldysh-Floquet and the Bloch-RWA results. This regime corresponds to weak coupling with $\lambda \ll 1$.  
%so-called weak-coupling regime in quantum optics, with $\mathcal{V}_0/\Omega \ll 1$. 
%In the weak-field ($E/E_0 \lesssim 0.2$) regime, as $\eta$ decreases
%below $\delta$, \textcolor{blue}{we achieve close agreement between the Keldysh-Floquet
%and the Bloch-RWA results, as expected in the coherent regime ($\eta \to 0$).} 
%This is because the Bloch-RWA theory works only in the coherent
%regime ($\eta \to 0$). 
On the other hand, for stronger fields ($E/E_0 \gtrsim 0.2$), 
%in the strong-field ($E/E_0 \gtrsim 0.2$) regime,
the Keldysh-Floquet result differs noticeably from the Bloch-RWA result, because $n$-photon excitation processes ($n \geq 2$) that becomes important at strong fields are ignored in the RWA~\cite{Remark3}, but are exactly captured in the Keldysh-Floquet approach.

%%%%%%%%%%%%%%%%%%%%%%%
%%%%%%%%%%%%%%%%%%%%%%%

%%%%%%%%%%%%%%%%%%%%%%%%
%                     Dynamical QAHE                      %
%%%%%%%%%%%%%%%%%%%%%%%%

%%%%%%%%%%%%%%%%%%%%%%%
\begin{figure}[t]
\centering
\includegraphics[width=0.45\textwidth]
{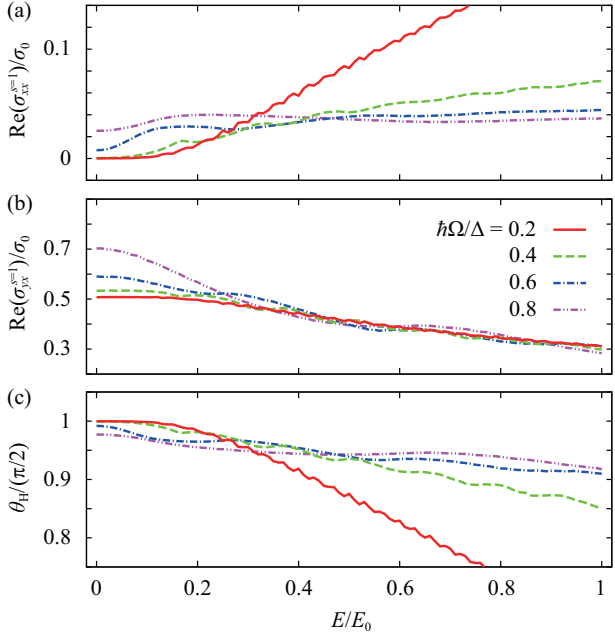} \\
\caption{
Dynamical breakdown of the one-half QAHE. Each panel shows (a) the dynamical longitudinal conductivity, (b) the dynamical Hall conductivity, and (c) the generalized Hall angle as a function of optical field strength $E$ for some values of optical frequency $\Omega$. Here, we used the parameters $k_B T / \Delta = 0.02$, $\eta / \Delta = 0.01$, and $\mathcal{E}_c / \Delta = 10$.}
\label{Figure2}
\end{figure}
%%%%%%%%%%%%%%%%%%%%%%%

%%%%%%%%%%%%%%%%%%%%%%%
% \begin{figure}[t]
% \centering
% \includegraphics[width=0.46\textwidth]
% {Figure3.eps} \\
% \caption{
% (a) Leftmost panel: Time-averaged local density of states (LDOS) for Floquet states with $\hbar\Omega / \Delta = 0.4$ and $E/E_0 = 0.5$ (red line); the equilibrium counterpart is depicted by the green line. Right panels: LDOS shifted by $n\hbar\Omega$ ($n\in\{1,2,3,4\}$). The shifted LDOS is overlapped with the original one within the optical energy window (blue line) to determine the weight function $\mathcal{Q}_n$. 
% (b)--(d) show the weight function $Q_n$, $W_n$, and their product $Q_n W_n$, respectively, as a function of optical field strength $E$ for $\hbar\Omega / \Delta = 0.4$. Unassigned parameters are the same as in Fig.~\ref{Figure2}.}
% \label{Figure3}
% \end{figure}
%%%%%%%%%%%%%%%%%%%%%%%

%%%%%%%%%%%%%%%%%%%%%%%
\begin{figure}[t]
\centering
\includegraphics[width=0.44\textwidth]
%{Figure4_1.eps} \\
{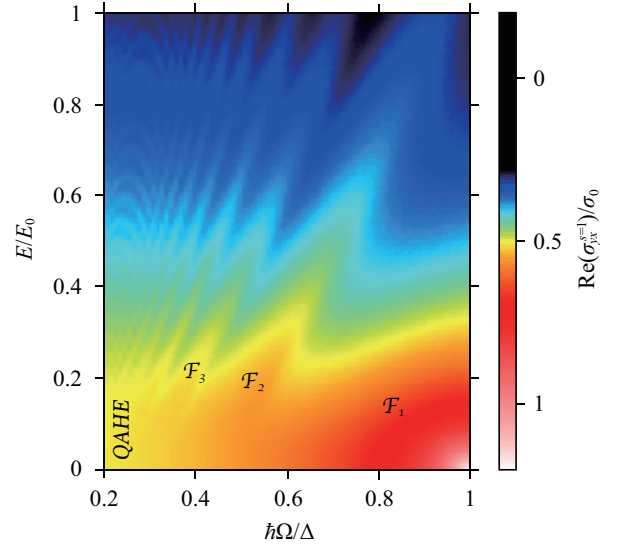} \\
\caption{
Phase diagram for the dynamical Hall conductivity as a function of the optical frequency $\Omega$ and the optical field strength $E$. The right panel indicates the color legend for the dynamical Hall conductivity. 
%For convenience, low-frequency data for $\hbar\Omega / \Delta < 0.2$ was excluded, because numerical computation in this regime needs relatively large Floquet space dimensions.
``QAHE" indicates the QAHE phase, and ``$\mathcal{F}_n$'' ($n\in\mathbb{N}$) the Floquet state with $n$-photon excitations. Unassigned parameters are the same as in  Fig.~\ref{Figure2}.}
%\label{Figure4}
\label{Figure3}
\end{figure}
%%%%%%%%%%%%%%%%%%%%%%%

\emph{Dynamical QAHE}.--- 
%\emph{Dynamical breakdown of QAHE}.--- 
%Our main interest is in the quantization regime, where the real part of the Hall conductivity is quantized and the real part of the longitudinal conductivity vanishes. 
Our main interest in this work is to investigate the robustness and the dynamical breakdown of the QAHE without restricting ourselves to the 
%linear response 
linear-response or near-resonance coherent regimes. Here, we focus our discussion on ${\rm Re}[\sigma_{xx}^{s=1}(E)]$ and ${\rm Re}[\sigma_{yx}^{s=1}(E)]$, which 
%entirely  capture the QAHE in the low-frequency regime.
are sufficient to capture the QAHE in the low-frequency regime. Figure~\ref{Figure2} shows %that 
%the collapse of the dynamical QAHE
the dynamical breakdown of the QAHE as a function of optical field strength for different frequencies. Both the longitudinal and Hall conductivities exhibit a consistent pattern underlain by the following two main features. 
%as described below. 
%We observe that the pattern of the dynamical breakdown is consistent in both the longitudinal and Hall conductivities. 
%(a) the dynamical longitudinal conductivity, (b) the dynamical Hall conductivity, and (c) the ratio between the two. 
%dynamical Hall angle. 
%In the following, we describe two main features underlying this
%pattern.
%
First, there is a clear signature for the robustness of the half-quantized Hall regime. In Fig.~\ref{Figure2}(b), for $\hbar\Omega / \Delta = 0.2$, low enough to capture the QAHE, the regime is robust up to a threshold optical field strength $E_{th} / E_0 \approx 0.12$. For $E < E_{th}$, the longitudinal counterpart is suppressed, and 
%the ratio between the dynamical Hall and the longitudinal conductivities 
the generalized Hall angle $\theta_H$, which is defined by $\tan^{-1}[{\rm Re}(\sigma_{yx}^{s=1}) / {\rm Re}(\sigma_{xx}^{s=1})]$, 
%hits 
attains the maximum value $\pi/2$ [see Fig.~\ref{Figure2}(a) and (c)]. 
Secondly, in the regime of $E > E_{th}$, 
%we notice that 
the QAHE 
%becomes unstable, 
dynamically collapses, and the conductivities exhibit an oscillatory behavior, which becomes more prominent with decreasing optical field frequency or increasing field strength.

In Fig.~\ref{Figure3}, we show the full dependence of the dynamical Hall conductivity on both the optical frequency and field strength. 
% dynamical Hall conductivity as a function of both the optical frequency and the optical field strength. 
The lower left corner of the plot labeled by 
%`QAHE'
``QAHE" corresponds to the low-field adiabatic regime in which the quantized Hall conductivity remains robust. The region with $E < E_{th} \approx 0.12 E_0$ and $\hbar \Omega \in (0, \Delta)$ corresponds to the 
%linear response
linear-response regime where the Kubo 
%formula 
theory is valid; whereas the region with $\hbar \Omega \approx \Delta$ and $E \in (0, E_0)$ is approximately captured by the 
%RWA-Bloch regime
Bloch-RWA theory when multiphoton processes are ignored. We observe that the oscillatory behavior is most prominent in the upper left corner of Fig.~\ref{Figure3}, where $\hbar\Omega/\Delta \approx 0.2$ and ${E/E_0} \lesssim 1$. This regime with low frequency and high electric field is characterized by a strong light-matter coupling, since $\lambda = ({E/E_0})/(\hbar\Omega/\Delta)^2 \gg 1$. Numerical calculations in this regime is challenging because of the large number of Floquet modes involved due to a small frequency. 
%interaction 
%$\mathcal{V}_0 \gg \mathcal{E}_{\bf k} \geq \Delta/2 \gg \Omega$, or $\lambda \gtrsim 1$. 
To understand the behavior in this regime, we treat the time-independent part of the Hamiltonian $\mathcal{H}_{\bf k}$ as a perturbation~\cite{Aguado}, valid when $\mathcal{V}_0 \gg \mathcal{E}_{\bf k} \geq \Delta/2$.   
%This allows us to treat the time-independent part of the Hamiltonian $\mathcal{H}_{\bf
%  k}$ as a perturbation \cite{Aguado} (see Supplementary Material for details). 
%The pattern of oscillations can be understood by considering the photon emission contribution $\tilde{\sigma}_{\rm emi}^{s=1}(E)$ to
%the longitudinal conductivity.  
%To first non-vanishing order in 
To linear order in $\mathcal{E}_{\bf k}/\mathcal{V}_0$,
%$\mathcal{H}_{\bf k}/\mathcal{V}_0$, 
we find that ${\rm Re} [\sigma_{yx}^{s=1}(E)] \propto \sum_{n\in\mathbb{N}} \mathcal{Q}_n \mathcal{W}_n$, 
where $\mathcal{Q}_n$ and
$\mathcal{W}_n$ are weight functions corresponding to 
%$n$-photon processes 
the Floquet state with $n$-photon excitations ($n\geq 1$), labeled by ``$\mathcal{F}_n$" in Fig. \ref{Figure3}:
\begin{align}
\mathcal{Q}_n
& = \int_{-\infty}^{\infty} d\tilde{\omega} \int_{-\infty}^{\infty} d\tilde{\omega}' 
\mathcal{P} \frac{1}{\tilde{\omega}' - \tilde{\omega} + \Omega} 
\rho(\tilde{\omega}) \rho(\tilde{\omega}'+n\Omega)
\nonumber\\
& ~~~ \times \big[ f_{\rm FD}(\tilde{\omega}) - f_{\rm FD}(\tilde{\omega}'+n\Omega) \big], \label{Qn}
%
% & = \int_{-\infty}^{\infty} d\hbar\tilde{\omega} \int_{-\infty}^{\infty} d\hbar\tilde{\omega}' \mathcal{P} \frac{1}{\hbar(\tilde{\omega}' - \tilde{\omega}) / \Delta} %\rho(\tilde{\omega}) \rho(\tilde{\omega}'+n\Omega)
% \nonumber\\
% & ~~~ \times \big[ f_{\rm FD}(\tilde{\omega}) - f_{\rm FD}(\tilde{\omega}'+n\Omega) \big],
\\
\mathcal{W}_n
& = 
\frac{E_0}{E} \mathbb{J}_{n-1}(2\lambda) [\mathbb{J}_{n}(2\lambda) - \mathbb{J}_{n-2}(2\lambda)] (1 - \delta_{n,1}/2),
\label{Wn}
%\mathcal{W}_n
%& = \frac{E_0}{E} \mathbb{J}_{n}\left( \frac{2E/E_0}{(\hbar\Omega/\Delta)^2} \right) \mathbb{J}_{n+1}\left( \frac{2E/E_0}{(\hbar\Omega/\Delta)^2} \right), \label{Wn}
\end{align}
where $\mathcal{P}$ stands for the principle value integral, 
%$\rho(\tilde{\omega}) = - {\rm Im} \sum_{\alpha\in\{\uparrow,\downarrow\}}[\hat{G}_{\bf k}^{R}(\omega)]_{\alpha\alpha;nn} / \pi$
$\rho(\tilde{\omega}) = - {\rm Im} \sum_{\bf k} \sum_{\alpha\in\{\uparrow,\downarrow\}}[\hat{G}_{\bf k}^{R}(\omega)]_{\alpha\alpha;nn} / \pi$ is the time-averaged local density of states with $\tilde{\omega} = \omega + n\Omega$, 
%$\rho(\tilde{\omega}) = - {\rm Im}
%\sum_{\alpha\in\{\uparrow,\downarrow\}}[\hat{G}_{\bf
%  k}^{R}(\omega)]_{\alpha\alpha;nn} / \pi$ with $\tilde{\omega} =
%\omega + n\Omega$, 
and $\mathbb{J}_{n}(x)$ is the Bessel function of
the first kind (see Supplementary Material for details). We note that this result is closely connected to the tunneling current formula under \textit{A.C.} bias voltage in the Tien-Gordon theory~\cite{Tien63}; in fact, the Hall conductivity with Eqs.~(\ref{Qn})-(\ref{Wn}) is in the form of the Kramers-Kronig counterpart of the Tien-Gordon tunneling conductivity. 
%bears a remarkable similarity with the
%tunneling formula in the Tien-Gordon theory~\cite{Tien63} (see Supplementary Material for details)
%This result captures the oscillatory features in the dynamical breakdown regime of the QAHE. 
%Using this result, we can understand the 
%oscillatory dependence of the Hall conductivity on the electric field in the dynamical 
%breakdown regime of the QAHE. 
Our numerical results show that the field dependence of $\mathcal{Q}_n$ is insignificant. Instead, the main effect is captured in the asymptotic form of $\mathcal{W}_n$ for $\lambda \gg 1$,
%$\mathcal{W}_n 
%\approx ({(-1)^{n+1}}/{2\pi})
%\left[({\hbar\Omega/\Delta})/({E/E_0})\right]^2 \cos(4\lambda)$,}
%
%$E/E_0 \gg (\hbar\Omega/\Delta)^2$,}
%Using the asymptotic expansion of the Bessel function, we find 
\begin{align}
\mathcal{W}_n 
\approx 
\frac{1}{2\pi} \left[ (-1)^{n} + \frac{1}{2} \delta_{n,1} \right] 
\frac{E_0}{E}\frac{\cos(4\lambda)}{\lambda},
%\approx \frac{(-1)^{n+1}}{2\pi}
%\left(\frac{\hbar\Omega/\Delta}{E/E_0}\right)^2 \cos\left[
%  \frac{4E/E_0}{(\hbar\Omega/\Delta)^2} \right],
%+ \mathcal{O}\left(\frac{(\hbar\Omega/\Delta)^2}{E/E_0}\right),
\label{Weight_Wn_asymp}
\end{align}
from which we see that the Hall current oscillates with the optical field strength $E/E_0$ at a frequency $\sim (\Delta/\hbar\Omega)^2$. 
%The oscillation frequency $2ev/(\hbar\Omega^2)$ as a function of $E$ can be measured experimentally to provide information on the Dirac velocity $v$. 
%Moreover, 
The oscillation frequency is therefore a direct probe of the light-matter coupling $\lambda$, with more frequent oscillations characterizing a stronger coupling.
%The Bessel function dependence in the Hall conductivity Eq.~(\ref{Wn}) raises the possibility that the Hall current can be dynamically suppressed in parameter regimes when %${2E/E_0}/(\hbar\Omega/\Delta)^2 = eEv/(\hbar\Omega^2)$ is a zero of the Bessel function, similar to dynamical localization studied in semiconductor superlattices. However, %further work is needed to confirm this possibility, as the exact Hall current obtained numerically is not simply proportional to a single Bessel function but is a sum of terms %containing different orders of Bessel functions.} 
%Finally, we close by discussing the conditions for experimental observation. 
%To measure dynamical QAHE, here we provide estimates for the optical frequency and field strength corresponding to the Fig.~\ref{Figure4}. 

For bismuth-based 
%topological insulators, 
TIs with Dirac velocity 
%is 
$v \approx 5 \times 10^5\,\mathrm{ms}^{-1}$ and magnetically-induced gap $\Delta = 0.02-0.2$ $\mathrm{eV}$~\cite{Chen10, AndrewWray11, Chang13, Chang15, Kou15}, we estimate that the required optical frequency and field strength for observing coherent oscillations of the Hall conductivity are $\Omega < 30.39-303.9\,\mathrm{THz}$ and $E \lesssim 1.215-121.5\,\mathrm{MVm}^{-1}$, which are well within current experimental accessibility. 
%dynamical QAHE 
The dynamical Hall conductivity in the TI film can be measured indirectly through magneto-optical Faraday and Kerr rotations, or directly in a standard Hall  measurement geometry illuminated with the linear polarization of the normally incident light parallel to the length of the Hall bar. 
In addition to TIs, graphene or bilayer graphene doped with noble metal atoms offer an alternate class of systems with a band gap and associated topological Hall transport \cite{QAHE_Tse2, Qiao_VHE}.
%In addition to TIs, graphene or bilayer graphene mounted on hBN substrate offer an alternate class of systems with a band gap and associated topological Hall transport \cite{}. 
In this scenario, the valley degrees of freedom give rise to a quantized valley Hall conductivity. Illuminated by a strong optical field, dynamical valley Hall currents will be generated in the transverse direction, which can be measured in a nonlocal transport geometry~\cite{VH1,VH2,VH3}.

In summary, we have developed a theory for the dynamical quantum anomalous Hall effect driven by intense optical fields. Our theory addresses the question of the robustness of topological Hall quantization in the nonlinear electric field regime, and predicts a collapse of Hall quantization at high fields accompanied by coherent conductivity oscillations  as a function of optical field strength. Our work sheds light on the problem of the nonequilibrium dynamical response of topological phases under a strong optical field, and our findings should offer new insights in nonequilibrium topological states particularly in the low-frequency, adiabatic regime.

This work is supported from a startup fund of the University of Alabama.

%%%%%%%%%%%%%%%%%%%%%%%%

%%%%%%%%%%%%%%%%%%%%%%%%


\begin{thebibliography}{99}
\bibitem{Remark2} To compensate for the cutoff-dependent correction $[1 - \Delta/(2\mathcal{E}_c)]$ to the Hall conductivity \cite{Tse16}, in the final plots of our numerical results, we rescale $\sigma_{yx}^{s}$ by the constant $[1 - \Delta/(2\mathcal{E}_c)]^{-1}$ so that the Hall conductivity in the limit of $\Omega \to 0$ approaches the exact half quantized value. This rescaling is necessary since a large $\mathcal{E}_c (\gg \Delta)$, which would ensure the correction factor approaches $1$, is not practical in our 
%computationally intensive
intensive numerical calculations. In all of our results, we have ensured that $\sigma_{yx}^{s}$ at 
%$\omega \ll \Delta$
$\hbar\Omega \ll \Delta$ and $E \ll E_0$ has converged to the theoretical value $(\sigma_0/2)[1-\Delta/(2\mathcal{E}_c)]$ before rescaling.
\bibitem{Remark3} In the RWA, only $n=1$ single-photon processes are included. This is because higher order  $n \geq 2$ resonant transitions require virtual transitions to intermediate states, and the counter-rotating terms that correspond to such virtual transition processes are ignored.
\end{thebibliography}
\end{document}